# A New Form-Factor Method for the Analysis of Tissue Fluorescence

# V. Gavryushin

*Institute of Materials Science and Applied Research and Semiconductor Physics department, Vilnius University, Vilnius, Lithuania. E-mail:* vladimiras.gavriusinas@ff.vu.lt

**Abstract:**

The aim of this article is to present a developed method that decomposes the autofluorescence spectrum into the spectra of naturally occurring biochemical components of biotissue. It requires knowledge of detailed spectrum behaviour of different endogenous fluorophores. We have studied the main bio-markers in human tissue and proposed a simple modelling algorithm for their spectra shapes. The empirical method was tested theoretically by quantum-mechanical calculations of the spectra in the unharmonic Morse potential approach.

## INTRODUCTION

Now is abundantly clear that the developments of laser use in the therapy and diagnostics (e.g., photodynamic therapy, optical tomography, 3D-microscopy, etc.) had been achieved. There has been increasing interest recently in the field of fluorescence-based techniques for the characterization of human lesions in clinical oncology. Several groups are developed fluorescence spectroscopy for early detection of the premalignant lesions [1-9].

Fluorescence spectrum of the tissue may be attributed primarily to the superposition of the fluorescence from a variety of interacting biological molecules, some of them as the naturally occurring fluorophores. We suggest that spectra decomposition to its constituents has to be done first [1,2], as is usual in the molecular spectroscopy, then followed by statistical analysis of decomposed elements or the whole spectra to find any correlation with medical indications of the biomedical object under study [2,3,4].

The question, then, arises: which of fluorescent chemical components one has to consider for the reconstruction of fluorescent spectra and how much of each has to be included for a perfect fit to overall fluorescence. In addition to that, it is well known, that the fluorescence emission line-shapes of macromolecules are strongly asymmetric. Therefore one might want to have a good enough model approximating the line shapes of the fluorophores. This is an issue I am addressing in this paper.

We have studied here the technique of resolving the total fluorescence spectra of tissue into the main fluorescent components by curve fitting [2]. A spectra resolution as a set of overlapping endogenous lines was carried out for better understanding of biochemical changes in normal and malignant biomedical samples [5]. For data evaluation and reconstruction of the line shapes of the studied fluorophores an empirical asymmetric-Gaussian-model, firstly used in [2], was introduced.

Since empirical method needs to be tested theoretically, a quantum calculation of the molecular emission spectra was done in an unharmonic Morse potential approach in the discreet variable representation [10]. Having done this, one could compare both the calculated and empirical spectra. It was found that the simple approach of "truncated Gaussian" goes in good agreement with measured asymmetric fluorescence spectral shapes and therefore is good enough to use as an approximation for it.

## EXPERIMENTAL

Pulsed laser-induced-fluorescence studies of pathologically certified tissues of premalignant and benign lesions in the female genital tract (uterus glandular cervical squamous epithelium) were carried out, as described in more detail in Ref. [2]. Laser induced autofluorescence (AF) measurements in vitro were performed with pulsed laser (3rd harmonic of $Nd^{+3}$-glass laser: $\hbar\omega_{exc}$ = 3,51 eV, 70 ns of pulse duration, 1Hz - repetition rate) as excitation source [2]. Tissue fluorescence spectra were measured by a "time-gated" (registration gate of 5 ns width and 10 ns delay after an excitation pulse) computer controlled, hand-made spectrophotometer [2]. Sixty patients were included in the study. For every tissue sample, three measurements (in different locations) were carried out. Spectra were used only from samples certified by histopathology analysis as normal/malignant. The conditions of excitation and of the collection of emission light were the same for all measurements. Laser light was focused ($\approx 1$ mm$^2$) to the sample pressed between two quartz plates.

The normalized experimental autofluorescence spectra of all tested tissue samples of different levels of pathology are shown in Figure 1a, and demonstrates the variations of the spectra of a same tissue type from one patient to another. Spectral data set was evaluated by MathCAD data processing software were written script, analyses every spectrum as the composition of the same set of asymmetric components. The used spectral components correspond to the known fluorophores [1,2,9,11].

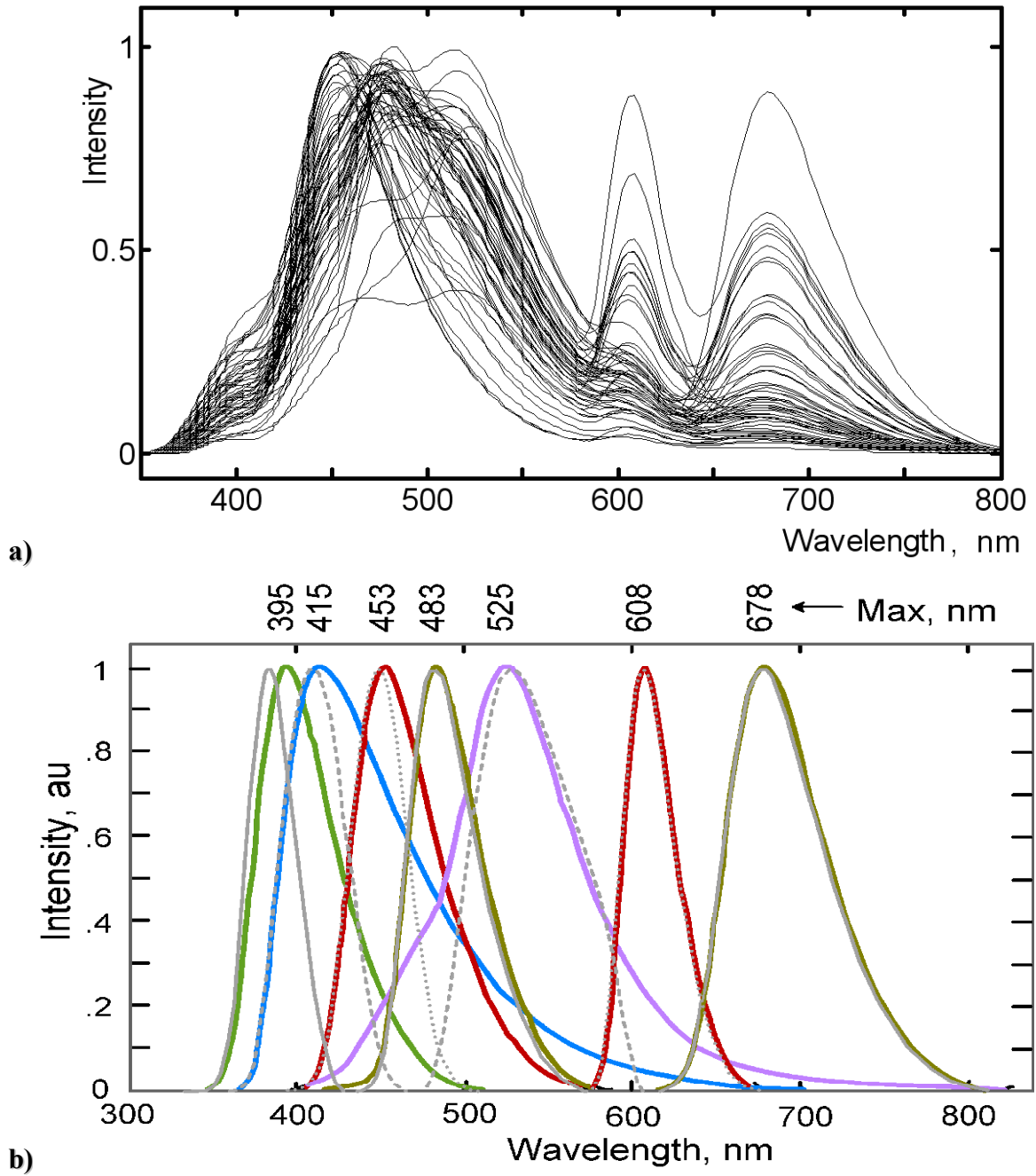

**a)**

**b)**

**Figure 1**

a) The set of the normalized autofluorescence spectra of the measured tissue samples of different levels of pathology, - from hyperplasia to cancer.

b) Fluorescence spectra of the main used endogenous fluorophores constructed by model (1)-(3): Collagen (395 nm), elastin (415 nm), NADH (453nm), unidentified (483 nm), caroten (525nm), porphyrins (HDP 608 & 678nm). Biomarker's spectra were derived from an experimental spectra (a). The spectral components, derived by multivariate curve resolution (Figure 2b), are shown by thin lines for comparison.

The intensity of emission of tested tissues had a large variation (about 2 orders) from patient to patient, also for tissues of the same patient, but for different excitation places. The spectral shape also display a large variation (Figure 1a), but more differ from patient to patient.

Therefore, the normalizing of a fluorescence spectra, as shown in Figure 1a, removes absolute intensity information reducing the interpatient and intrapatient variation in AF intensity.

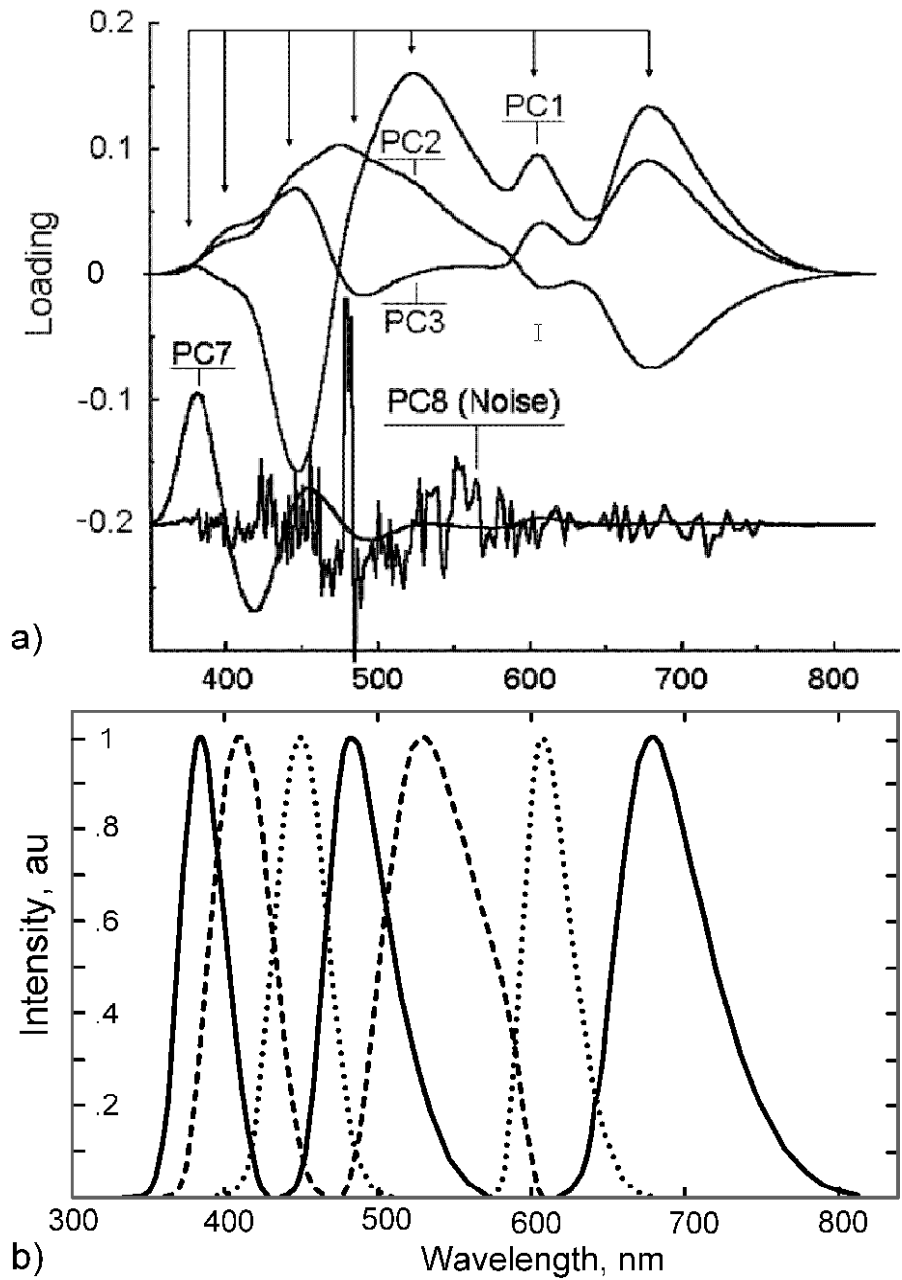

**Figure 2**

a) First three and shifted (-0.2) $7^{th}$ and $8^{th}$ principal components - factors considered as basis spectra characterizing main spectral changes in the data set (Figure 1a). The $8^{th}$ and further principal components describe only noise. These last principal components are discarded and 7 principal components are left for further analysis.

b) Spectra derived by multivariate curve resolution.

Extensive curve analysis was carried out in order to understand the changes in tissue biochemistry based on the spectral AF profiles. An effort has been made to estimate by multivariate curve resolution the minimal enough number of the tissue fluorophores forming AF spectra. Principal component analysis (PCA) [7,12] was used to derive principal components - basis spectra parts responsible for the spectral variation [3,4]. Derived principal components were tested with the Malinowski's indicator function [13] to select statistically significant principal components (factors). In Figure 2a, some principal components are shown. Remaining principal components describes only noise (8th and further, Figure 2a) and they are discarded from further analysis. Only seven factors were found to be statistically significant.

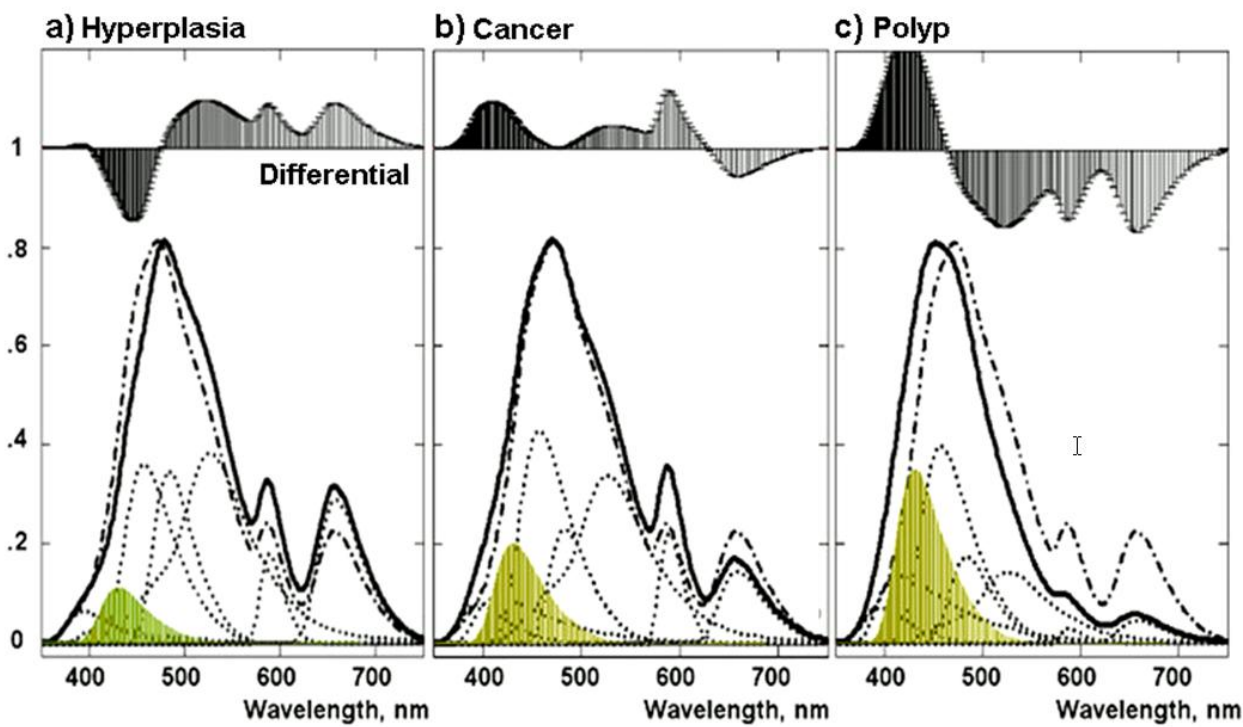

**Figure 3**

The comparison of averaged fluorescence spectra of samples from different histology groups (solid line) with the full average spectrum of all samples (broken line) [2]. Dotted lines show the main components forming the spectra (from right to left): HDP (678 nm); HDP (608 nm); flavin-caroten (525 nm); unidentified (483 nm); NADH (453 nm); neopterin (430 nm, darkened); elastin (415 nm); collagen (395 nm). On the top, - the difference spectra between averaged spectra of desease groups (solid) and full averaged spectrum (broken) are shown.

However, this factors are not yet directly related to any real chemical spectra (some of them, for example, are negative, Figure 2a). Non-oblique factor rotation transforms them to physically meaningful spectral profiles, then, alternating least squares MatLab's algorithm optimized the recovered profiles (Figure 2b) [4]. The factors, were used as initial estimates and non-negativity, unimodality and linear additivity of spectra, were assumed as chemical constraints. In Figure 2b, resolved spectral profiles by multivariate curve resolution are shown. Resolved spectra can be compared to known fluorophores spectra (Figure 1b) used in our previous work [2]. Obtained pure spectra can be used to identify chemical species using library spectra [6].

The evaluation of experimental spectra shows that autofluorescence requires at least 7 components to be fully accounted for all spectral changes [3]. Figure 1b shows the fluorescence spectra of used endogenous fluorophores as biomarkers for the spectral analysis. The known data for endogenous fluorophores are quite different [1,9]. Data evaluation [2,11] was made, firstly, by reconstruction of the biotissue spectra as the sum of the same set of line-shapes of endogenous fluorophores (Figure 1b): collagen, elastin, carotene, neopterin [11], NADH, FAD (nicotinamide and flavin adenine dinucleotides), porphyrins [14-18]. Further, an analysis was based on the difference of the tissue spectra from an averaged spectrum of healthy tissues.

Endometrial tissues samples were biopsied, classified by routine histopathology, and grouped as Normal, Hyperplastic (HP) or Cancerous (CN). This samples categories of different diagnosis were analyzed [2,3,11]. Mean-scaling was performed by calculating the **mean spectrum** (broken line spectra in Figure 3) for all more healthy patients and subtracting it from each patient spectrum and from an averaged spectra for patients grouped by same diagnosis. However, unlike normalization, mean-scaling displays the differences (dashed spectra on the top of Figure 3) in autofluorescence spectra with respect to the artificial healthy tissue averaged spectrum. Therefore, this method maximally enhances the differences in autofluorescence spectra between tissue categories when spectra are acquired from non-diseased and diseased sites from each patient. As an example of aforementioned spectra-processing, results are shown in Figure 3 [2,3]. A difference factors between averaged spectra of the groups (solid lines) and the healthy tissue averaged spectrum (broken lines) are shown on the top of the each spectrum. As we can see, the neopterin presence in cancer (b) and polyp (c) stages of the tissue is about twice more evident, and more, has opposite changes comparing to hyperplasia (a). Opposite changes in the difference factor stand for enhancement of its presence, in relationship to the mean composition, while in hyperplastic case (a), its presence is weakened [2,11].

## FORM-FACTOR FOR SPECTRA MODELING

For the reconstruction of the spectra of fluorophores, as bio-markers, the normalized line-shape function $S_i(\hbar\omega)$ can be calculated as an asymmetric-Gaussian [2, 11]:

$$S_i(\hbar\omega) = \frac{1}{\int_1^5 S_{i0}(\hbar\omega)} G_i(\hbar\omega) \cdot D_{cut_i}(\hbar\omega), \qquad (1)$$

$$G_i(\hbar\omega) = \frac{1}{\sqrt{\pi}\Delta_i} \exp[-\left(\frac{\hbar\omega - E_{0_i}}{\Delta_i}\right)^2], \qquad (2)$$

$$D_{cut_i}(\hbar\omega) = \exp[-\left(\frac{\hbar\omega - E_{0_i} - \gamma_i\Delta_i}{\gamma_i\Delta_i}\right)^{g_i}]. \qquad (3)$$

Here $G_i(\hbar\omega)$ is an usual symmetrical Gaussian function of a half-width $\Delta_i$ and the maxima energy

$E_0$; $D_{Cut}(\hbar\omega)$ is a cutting factor-function (of a $g$-rank) deforming the initial Gaussian; $\gamma$ is a width factor of the cutting function; $\int_1^5 S_0(\hbar\omega) = 0.445$ is the integral, normalizing spectral function $S_i(\hbar\omega)$ area to 1. Rather like attempts for the molecular line-shape approximations can be found also in [19,20].

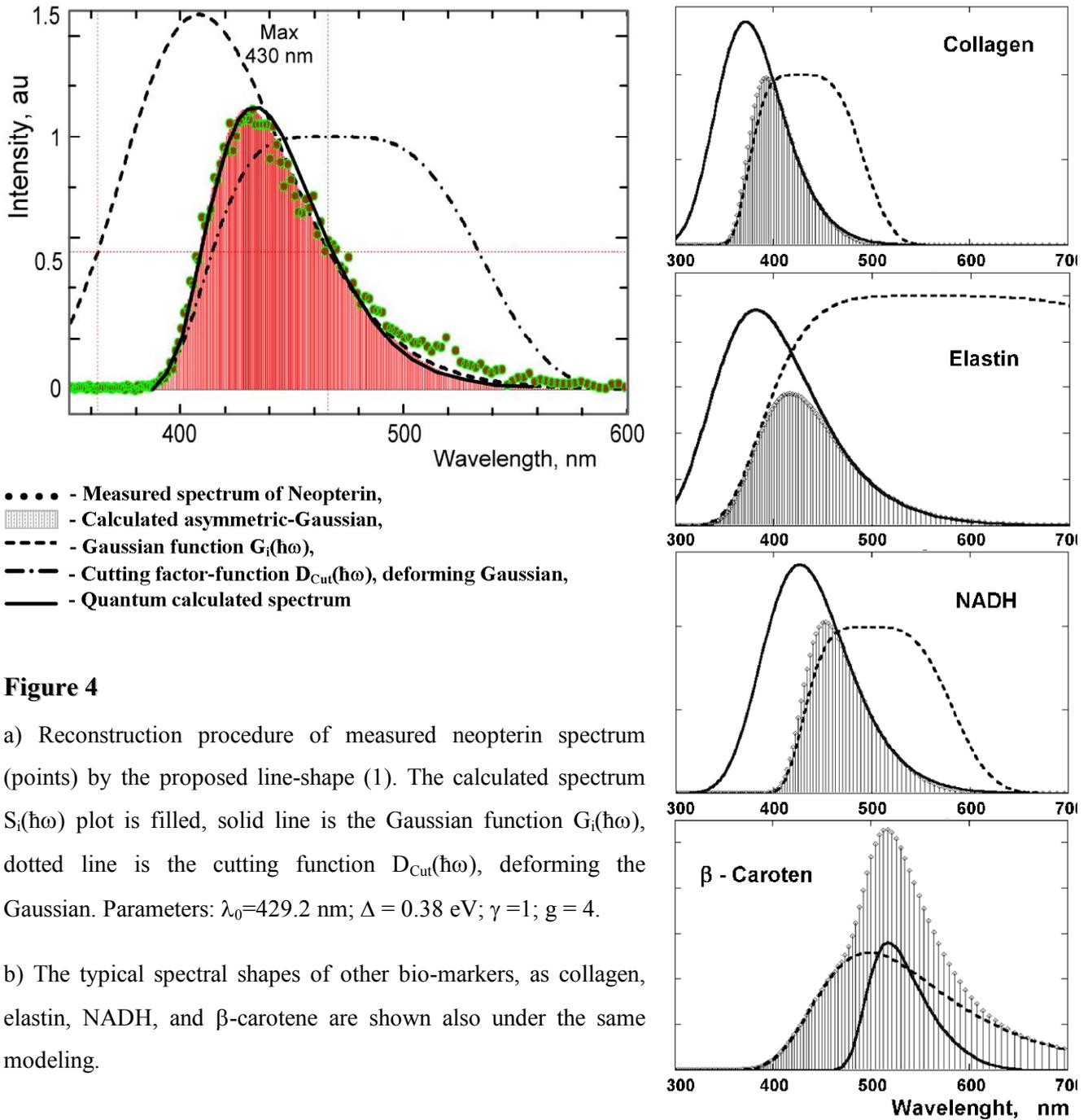

• • • • - Measured spectrum of Neopterin,
         - Calculated asymmetric-Gaussian,
- - - - - Gaussian function $G_i(\hbar\omega)$,
- · - · - Cutting factor-function $D_{Cut}(\hbar\omega)$, deforming Gaussian,
         - Quantum calculated spectrum

**Figure 4**

a) Reconstruction procedure of measured neopterin spectrum (points) by the proposed line-shape (1). The calculated spectrum $S_i(\hbar\omega)$ plot is filled, solid line is the Gaussian function $G_i(\hbar\omega)$, dotted line is the cutting function $D_{Cut}(\hbar\omega)$, deforming the Gaussian. Parameters: $\lambda_0 = 429.2$ nm; $\Delta = 0.38$ eV; $\gamma = 1$; $g = 4$.

b) The typical spectral shapes of other bio-markers, as collagen, elastin, NADH, and β-carotene are shown also under the same modeling.

As an example, such reconstruction of our measured [8,11] spectrum of a neopterin (points) together with factorizing components (2) and (3) are shown on Figure 4. We show also the spectrum (by broken line over points) obtained from the below following quantum calculations (same as one in Figure 7). The good correspondence between experimental, empirical and calculated below spectra allow us to state that a simple asymmetric-Gaussian model (1) - (3) as line-shape approximation is good enough to be used for real spectral analysis. Usually it is enough to change only two parameters in (1) - (3): the spectral position $\lambda_0$ and the spectral width $\Delta_i$. Therefore, we have the simple 2-parameter approximation for the tabulation of the spectral components as bio-markers. The typical spectral shapes of other biomarkers, such as collagen, elastin, NADH, and carotene are also modeled similarly and are shown in Figure 4b. It is noteworthy that all spectral components in Figure 1b and Figure 3 were calculated also by proposed model.

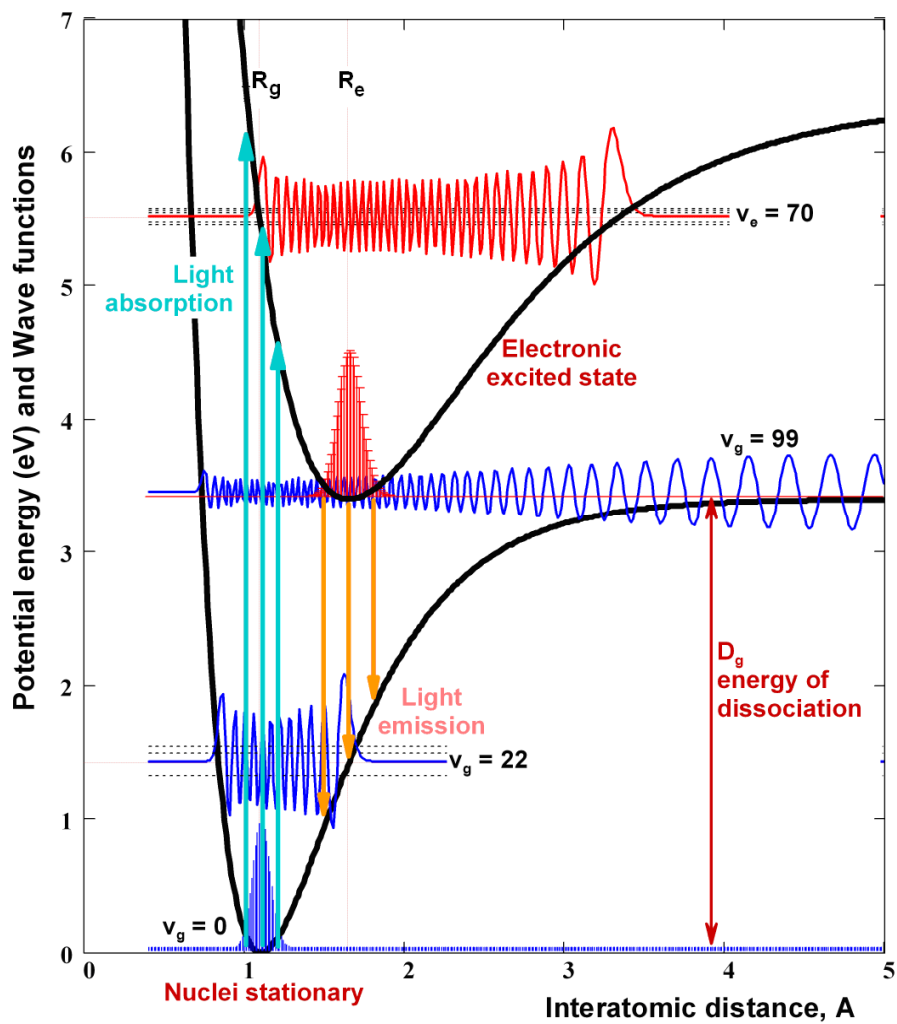

**Figure 5**

Morse potentials and some of the calculated eigenstates and wavefunctions for the ground *g* and excited *e* molecular states. Arrows show the transitions for light emission and absorption.

## QUANTUM CALCULATIONS OF LINE SHAPE

For theoretical arguments of above simple spectra modeling, we carried out the quantum calculations of stationary Schrödinger equation in an unharmonic Morse potential approach. The vibrational wavefunctions were obtained within discrete variable representation (DVR) method [21]. The DVR is a grid-point representation in which the matrix elements of the potential energy operator V(r) is approximated as diagonal and the kinetic energy as a sum of one-dimensional matrices [22, 23]. The formalism is permitting the line shape calculations starting only from a given interaction potential. Some of the calculated wavefunctions and eigenvalues are shown in Figure 5. We have calculated the spectra for emission and absorption (Figure 7) and compared them with an empirical approximation (1) presented above.

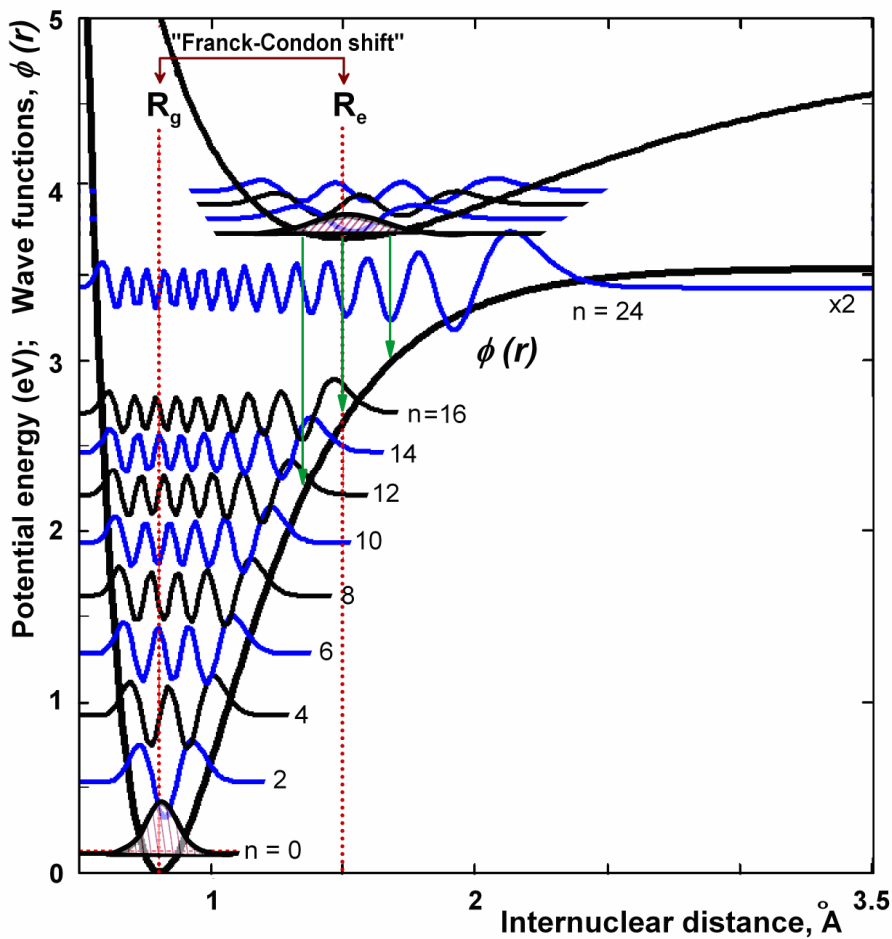

**Figure 6**

Potential curves and some of the calculated wavefunctions for the ground **g** and excited **e** molecular states calculated for some value of Franck-Condon (FC) shift. Arrows - light emission transitions.

### Morse potential approximation

The Morse potential is actually a good representation for the molecular potential energy and gives as the asymmetrically shaped emission and absorption lines. The Morse potential is

$$V(R) = D_e(1 - e^{-\beta(R - R_e)})^2 , \qquad \beta = \pi v_e \sqrt{\frac{2\mu}{D_e}} \qquad (4)$$

$v_e$ is the vibrational constant and **μ** is the reduced mass of molecule. Repulsive forces make the potential steeper than parabolic when the bond length is compressed and less steep when the bond length is expanded (Figure 5).

**Schrödinger Equation solution in Discreet Variable Representation (DVR)**

We select DVR method [24, 25] for our MathCAD calculations since it avoid having to evaluate integrals in order to obtain the Hamiltonian matrix and since an energy truncation procedure allows the DVR grid points to be adapted naturally to the shape of any given potential energy surface. The primary feature is that DVR gives an extremely simple kinetic energy matrix $\hat{T}_{i,k} = \hbar^2 k_{i,k}^2 / 2\mu$ with conditional formulation:

$$\hat{T}_{i,k} = if\left[ i \neq k, g\frac{2(-1)^{i-k}}{(i-k)^2}, g\frac{\pi^2}{3} \right] \tag{5}$$

We use the energetically weighted grid parameter **g** depending on a number of calculus points $i_{min}$:

$$g = \frac{\hbar^2}{2\mu}\left(\frac{1}{dr}\right)^2 \qquad dr = \frac{r_{max} - r_{min}}{i_{min}} \tag{6}$$

Here $r_{max}$ and $r_{min}$ defines the range variables for the bond length r. The diagonal matrix of the potential energy $\hat{V}_{i,k}$ for Morse potential (4) is as:

$$\hat{V}_{i,k} = if(i \neq k, 0, else, V_i) \tag{7}$$

Then matrix for the full Hamiltonian (in Hartrees) will be:

$$\hat{H}_{i,k} = \hat{T}_{i,k} + \hat{V}_{i,k} \tag{8}$$

The eigenvalues (stationary states) as the solutions of the Schrödinger equation $\hat{H}\psi_i(r) = E\psi_i(r)$ then are sorted:

$$e = sort(eigenvals(\hat{H})). \tag{9}$$

Wavefunctions for every of eigenvalues $n$=0,..,99 are got by simple MathCAD procedure:

$$\psi^{\langle n\rangle} = eigenvec(\hat{H}, e_n). \tag{10}$$

Finally, wavefunctions was normalized in atomic units:

$$\psi = \frac{\psi}{\sqrt{N}}, \qquad N = \frac{1}{a_0}\left[ \sum_i \psi_i^{\langle n\rangle}\psi_i^{\langle n\rangle}dr \right]. \tag{11}$$

Some of calculated wavefunctions and energy states are shown in Figure 5. An important feature is the increase of density of states of the vibrational energy levels with larger quantum number. The states over the dissociation energy $D_g$ are unbound and delocalized as you can see for the state $v_g = 99$ in Figure 5. Moreover, it is seen that in some cases the emission from an excited state can be quenched resonantly by unbound ground states.

**Electronic Transitions and Franck-Condon Factors**

The emission and absorption spectra of molecules was defined by the dipole matrix elements of optical transitions as the integrals $\langle \parallel \rangle$ over the vibrational coordinates:

$$\left\langle \Psi_{el.vibr}^{final} \left| \hat{\mu} \right| \Psi_{el.vibr}^{init} \right\rangle = \left\langle \Psi_{electr}^{final} \left| \hat{\mu} \right| \Psi_{electr}^{init} \right\rangle \times \left\langle \Psi_{vibr}^{final} \mid \Psi_{vibr}^{init} \right\rangle, \tag{12}$$

which has the form of the dipole integral of electronic transitions multiplied by the "overlap integral" between the initial and final vibrational wavefunctions (Figure 6). It is clear that, due to the asymmetry of molecular potentials, the emission spectra will be wider than the absorption spectra.

The squares of overlap integrals $\left\langle \Psi_{vibr}^{final} \mid \Psi_{vibr}^{init} \right\rangle$ inside the transition rate expression are called as "Franck-Condon factors" (FCF). Their magnitude plays the main role determining the relative intensities of vibrational "bands" within a particular spectrum of electronic transitions. Therefore, FCF determines the shapes of spectral lines. The overlap of wavefunctions (Figure 6) in forms $\langle g_0|e_i \rangle$ and $\langle e_0|g_i \rangle$, determining the absorption and emission probabilities, are calculated and their spectra are shown in Figure 7 for different values of the shifts of potential minima: $dr = R_e - R_g$. If the difference in potential curves minima $dr$ is growing then the spectra became wider.

Fluorescence spectra of bio-molecules, as have been shown on simple quantum-theoretical background, are of asymmetric profiles. One of the spectra from Figure 7 is drawn on Figure 4 (broken line over points) over the measured neopterin spectrum (points). The good agreement between quantum-calculated, measured, and empirical approximated (1) spectra are evident.

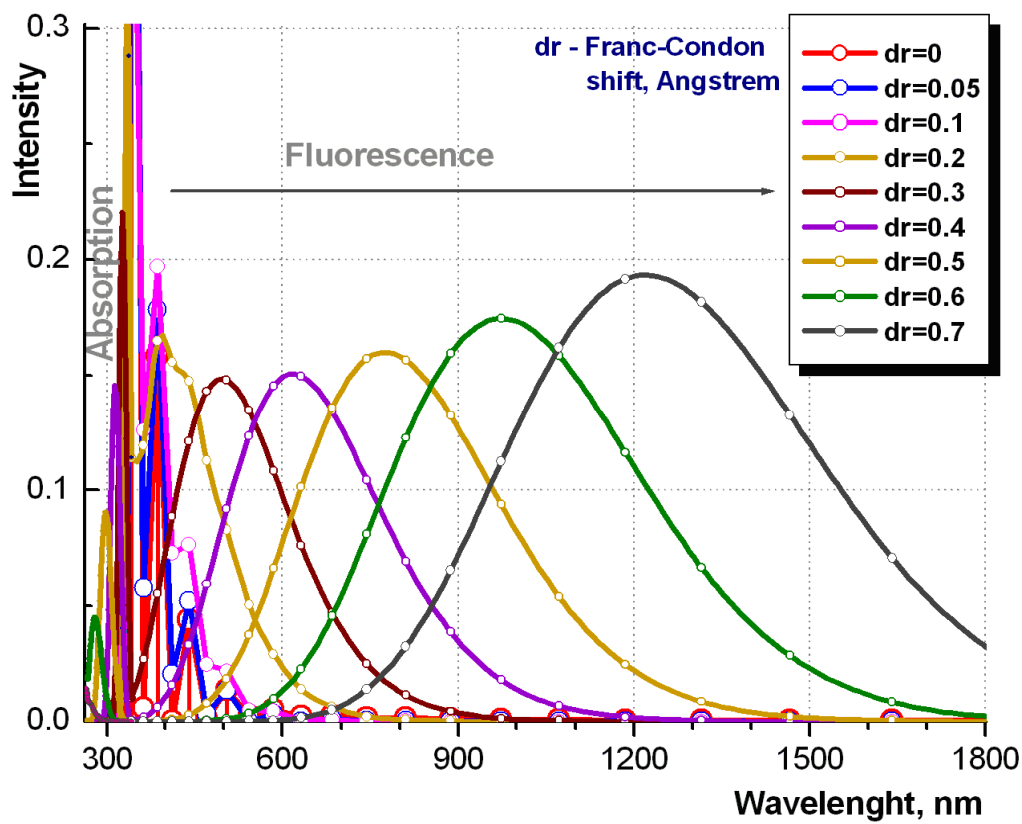

**Figure 7**

Fluorescence and absorption spectra calculated in Morse potential approach for different values of Franck-Condon shift" **dr** (Å).

## CONCLUSION

A method was discussed for the biotissue autofluorescence spectra decomposition into its constituent's fluorescence spectra. The decomposed spectra correspond to naturally, within the object occurring fluorescent substances which, if its concentrations correlate with medical indications, serves as a biomarker for disease diagnostics.

Fluorescence line-shape, on its own, was shown to be explained by two-parameter-only asymmetric-Gaussian model. To prove this, a quantum calculation of molecular emission spectra using Morse potential approach was employed. The comparison of the both calculated and empirical spectra with experimental one shows that simple spectra modeling [2,11] is useful for the description of the biomedical object spectra.

## ACKNOWLEDGMENTS

The author wish to thank colleagues E. Auksorius for his skilled support and data statistical treatments and MD. A. Vaitkuviene for the expert support on medicine aspects.